\documentclass[%
preprint,
showpacs,
showkeys,
 amsmath,amssymb,
 aps,
pra,
]{revtex4-1}

\usepackage{graphicx}
\usepackage{dcolumn}
\usepackage{bm}
\usepackage{hyperref}

\begin{document}

\preprint{APS/123-QED}

\title{Approximating the entire spectrum of nonequilibrium steady state distributions using relative entropy: An application to thermal conduction}

\author{Puneet Kumar Patra}
\affiliation{%
 Advanced Technology Development Center, Indian Institute of Technology Kharagpur, India, 721302 
}%
\author{Marc Mel\'{e}ndez}%
 \email{mmelendez@fisfun.uned.es}
\affiliation{%
Dpto. F\'{i}sica Fundamental, UNED, Madrid, Spain, 28040 
}%
\author{Baidurya Bhattacharya}%
 \email{Corresponding Author: baidurya@civil.iitkgp.ernet.in}
\affiliation{%
Department of Civil Engineering, Indian Institute of Technology Kharagpur, India, 721302 
}%

\date{\today}

\begin{abstract}
We show that distribution functions of nonequilibrium steady states (NESS) evolving under a slowly varying protocol can be accurately obtained from limited data and the closest known detailed state of the system. In this manner, one needs to perform only a few detailed experiments to obtain the nonequilibrium distribution function for the entire gamut of nonlinearity. We achieve this by maximizing the relative entropy functional (MaxRent), which is proportional to the Kullback-Leibler distance from a known density function, subject to constraints supplied by the problem definition and new measurements. MaxRent is thus superior to the principle of maximum entropy (MaxEnt), which maximizes Shannon's informational entropy for estimating distributions  but lacks the ability of incorporating additional prior information. The MaxRent principle is illustrated using a toy model of $\phi^4$ thermal conduction consisting of a single lattice point. An external protocol controlled position-dependent temperature field drives the system out of equilibrium. Two different thermostatting schemes are employed: the Hoover-Holian deterministic thermostat (which produces multifractal dynamics under strong nonlinearity) and the Langevin stochastic thermostat (which produces phase space-filling dynamics). Out of the 80 possible states produced by the protocol, we assume that 4 states are known to us in detail, one of which is used as input into MaxRent at a time.  We find that MaxRent accurately approximates the phase space density functions at all values of the protocol even when the known distribution is far away. MaxEnt, however, is unable to capture the fine details of the phase space distribution functions.  We expect this method to be useful in other external protocol driven nonequilibrium cases as well, making it unnecessary to perform detailed experiments for all the values of the protocol when one wishes to obtain approximate distributions.

\end{abstract}

\maketitle

\section{\label{section1}Introduction \protect\\}
One of the challenges of non-equilibrium thermodynamics lies in the unified statistical description of nonequilibrium systems \cite{jou_eit}. These systems are usually characterized by the presence of non-vanishing currents (like heat, mass etc.) due to the imposed thermodynamic forces (like temperature gradient, concentration gradient etc.). A sound theoretical understanding of these nonequilibrium processes is vital to gain insights into several important physical and biological processes that span the length scale of a few atoms to billion-atom systems. The nonequilibrium systems may be classified into - systems near the equilibrium regime (local thermodynamic equilibrium) and systems far-from-equilibrium. Unlike in the equilibrium and near-equilibrium scenario \cite{de_groot_mazur_book,jou_book_2}, no unified theoretical framework exists for the far-from-equilibrium cases \cite{vavruch_et_al}. The problem becomes all the more difficult since there is no consensus on the variables necessary to describe the nonequilibrium states \cite{temperature_review, goldstein_boltz_entropy}. As a result, limited statistical descriptions of NESS are usually obtained through either numerical simulations \cite{corbet_maxent2, corbet_maxent,miller_larson,nisbet_gurney} or experiments \cite{ghosh,seitaridou_et_al,david_et_al}. Experimental techniques have matured enough to perform nanoscale experiments under externally controlled protocols, like the diffusion of a few particles in response to a variation in the concentration gradient, and single particle reaction dynamics by varying the energy barrier between two successive equilibrium positions controlled by laser traps. 

Usually, the limited information from such experiments is used to construct the probability distribution of the relevant variable through Jaynes’ principle of maximum entropy (MaxEnt)\cite{jaynes_maxent1,jaynes_maxent2,shore_johnson, skilling_2, presse_et_al, bamberg_book, mead_moments, grendar_maxent, emch_book}. Given a set of macroscopic observables: $\langle F_1 \rangle, \langle F_2 \rangle, ..., \langle F_N \rangle$ corresponding to the phase functions $F_1\left(\Gamma\right), F_2\left(\Gamma\right), ..., F_N\left(\Gamma\right)$, MaxEnt finds the least biased probability distribution $\bar{\rho}\left(\Gamma\right)$ by maximizing the Shannon entropy functional subjected to the constraints imposed by the set of macroscopic observables. For continuous states, the Shannon entropy augmented with the constraint equations can be written as:
\begin{equation}
H \equiv -k_B \int\limits_{\Omega}\rho(\Gamma) \log(\rho(\Gamma))d\Gamma - \sum\limits_{j=1}^{N}\lambda_j \left[\int\limits_{\Omega}F_j(\Gamma)\rho(\Gamma)d\Gamma - \langle F_j \rangle \right].
\label{maxent_entropy}
\end{equation}
In (\ref{maxent_entropy}), $\lambda_j$ represents the Lagrange multiplier corresponding to the observable $\langle F_j \rangle$. The approximated probability distribution $\bar{\rho}(\Gamma)$ is obtained by taking variation of (\ref{maxent_entropy}) with respect to the unknown distribution function $\rho(\Gamma)$,
\begin{equation}
\bar{\rho}\left( \Gamma \right) = \dfrac{1}{Z} \exp \left[ - \sum\limits_j \lambda_j F_j(\Gamma) \right].
\label{maxent_approx_dist}
\end{equation}
In (\ref{maxent_approx_dist}), $Z$ is the generalized partition function which arises from the normalization of the probability density function and is given by \cite{jou_book}
\begin{equation}
Z = \int\limits_\Omega \exp \left[ - \sum\limits_j \lambda_j F_j (\Gamma) \right].
\label{parition_function}
\end{equation}
The macroscopic observables, $\langle F_j \rangle$, are related to the partition function through the relationship  $\langle F_j \rangle = \partial \log Z / \partial \lambda_j$. To complete the approximated distribution, one needs to find the solution of the unknown Lagrange multipliers. This procedure is similar to the well-known problem of moments \cite{agmon, moments_solution}. For the equilibrium scenario, the Lagrange multipliers can be interpreted in physical terms without any numerical computation. For example, when the external constraint is chosen as the average internal energy   $\langle U \rangle$ of the system, the corresponding Lagrange multiplier can be identified as $\lambda = 1/k_BT$ and one can recover the canonical distribution function. The different equilibrium distribution functions can be obtained by judiciously choosing the constraints. MaxEnt has been successfully employed towards constructing NESS distribution functions for steady state thermal conduction \cite{corbet_maxent, corbet_maxent2, miller_larson, nisbet_gurney}. The flexibility offered by the MaxEnt formalism has rendered it useful in situations beyond statistical physics \cite{banavar_maxrent_application, skilling_3, skilling_maxrent}. 

However, MaxEnt formalism suffers from the inability of incorporating any information other than that of constraints \cite{shore_3}. Let us say that we have the  complete distribution function of the system at some value of the external protocol for the experiments discussed before, and suppose that the protocol is altered resulting in a new steady state for which very limited information is obtained by performing a not-so-detailed experiment. If we now wish to use MaxEnt to estimate the current NESS distribution function, we will not be able to utilize the detailed information of the previous steady state mentioned above. Rather, we will have to confine ourselves to using only the limited experimental results obtained in the current state and all previous knowledge on the system (obtained possibly at a significant cost) will go unutilized.

In this work, we show that by maximizing the relative entropy functional (MaxRent) one eliminates the problem explained in the previous paragraph. One need not perform detailed experiments for all the values of the protocol. With limited data of the system at all states, and a detailed probability description of one or a few of these states, it is possible to construct the nonequilibrium distribution functions for the entire range of nonlinearity. We demonstrate the efficacy of MaxRent using a toy model of thermal conduction: a $\phi^4$ thermal conducting model consisting of a single lattice point brought out of equilibrium by a position-dependent temperature field, whose strength evolves slowly through external control. We use two different schemes to impose the temperature field: the Hoover-Holian thermostat \cite{hh_thermostat}  and the Langevin thermostat \cite{langevin_thermostat}. Using the complete statistics at an earlier steady state and a few simple constraints at the current state that have direct physical meaning out of equilibrium, we demonstrate that, as the protocol evolves, MaxRent is able to approximate the true distribution function, $\rho_t(\Gamma)$, much more accurately than MaxEnt. The paper is organized as follows: in the next section we detail the toy model of thermal conduction followed by the details of MaxRent formalism. Next we provide numerical evidence for the efficacy of MaxRent.

\section{\label{section2}The Thermal Conduction Model}

The toy model employed in this study is a $\phi^4$ model \cite{aoki_phi4, aoki_phi4_2} comprising a single lattice point and governed by the potential
\begin{equation}
U\left(x\right) = \dfrac{1}{4}x^4.
\label{phi4_potential}
\end{equation}
Here, $x$ denotes the displacement from the equilibrium position (the origin). The mass of this single quartic oscillator is taken to be one. This model is chosen because it shows all the qualitative features of a system with many degrees of freedom and yet, is simple to analyze. The system is brought out of equilibrium by imposing a position-dependent temperature field \cite{posch_hoover, hoover_multifractal, hoover_kum, sprott_et_al}:
\begin{equation}
T\left(x\right) = 1 + \epsilon \tanh(x).
\label{temperature_field}
\end{equation}
The parameter $\epsilon$ acts as the external protocol and defines the strength of nonlinearity. A value of zero indicates equilibrium (canonical) while a positive value indicates a nonequilibrium state. A far-from-equilibrium scenario occurs when $\epsilon$ is large (but less than 1). For simplicity we have taken Boltzmann constant to be unity. The  imposed temperature field would cause the heat flow to occur in a direction opposite to the temperature gradient. In this work, $\epsilon$ is varied from 0 to 0.80 in steps of 0.01. The temperature is imposed on the system through the Langevin stochastic thermostat \cite{langevin_thermostat} (LT) and the Hoover-Holian deterministic thermostat \cite{hh_thermostat} (HH). The LT equations of motion for this case are
\begin{equation}
\begin{array}{cc}
\dot{x} = p, & \dot{p} = -x^3 - \phi p +\psi,
\end{array}
\label{lt_eom}
\end{equation}
where the damping constant $\phi$ and the stochastic force $\psi$ in (\ref{lt_eom}) are related through:
\begin{equation}
\begin{array}{cc}
\langle \psi(t) \rangle = 0, & \langle \psi(t) \psi(t') \rangle = 2\phi T(x) \delta(t-t').
\end{array}
\label{lt_relations}
\end{equation}
In the present study $\phi = 0.5$. The equations of motion for the HH thermostat are:
\begin{equation}
\begin{array}{cccc}
\dot{x} = p, & \dot{p} = -x^3 - \eta p - \xi p^3, & \dot{\eta} = p^2 - T(x), & \dot{\xi} = p^4 - 3T(x)p^2.
\end{array}
\label{hh_eom}
\end{equation}
The variables $\eta$ and $\xi$ represent reservoir variables that control the second and the fourth moments of velocity, respectively. The masses corresponding to the reservoir variables have been assumed to be unity. The equations of motion (\ref{lt_eom}) and (\ref{hh_eom}) are integrated using the classical fourth order Runge-Kutta method for 1 billion time steps with $\Delta t = 0.001$ for every realization of  $\epsilon$.

\begin{figure*}
\includegraphics[scale=0.40]{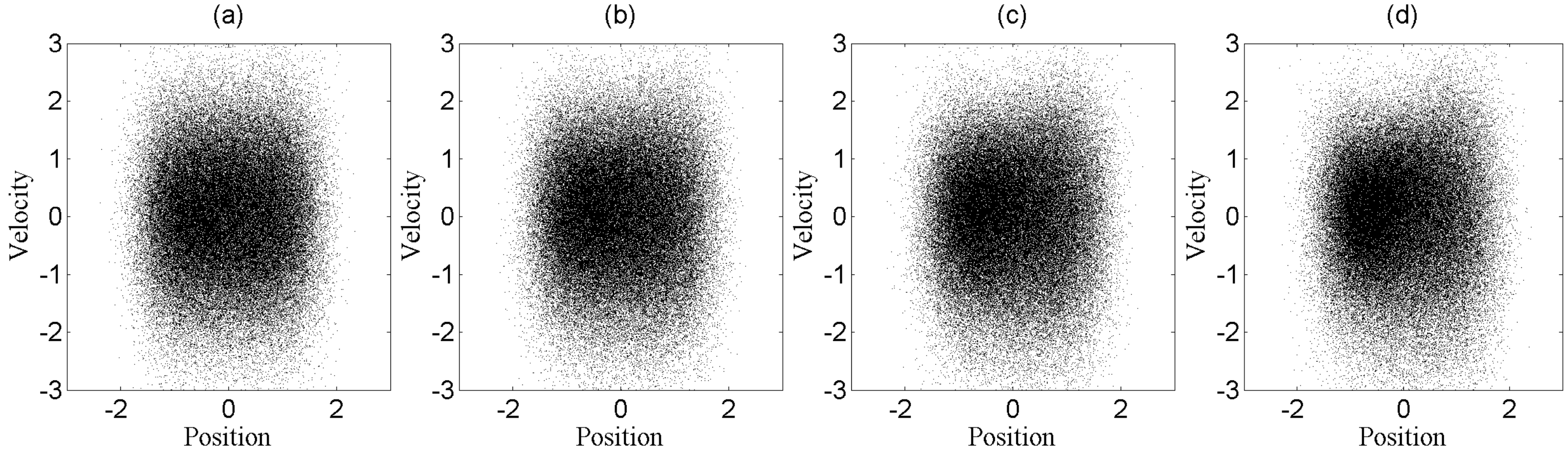}
\caption{\label{fig:FIGURE1} Phase space trajectory for Langevin dynamics. (a) corresponds to $\epsilon = 0.25$; (b) corresponds to $\epsilon = 0.50$; (c) corresponds to $\epsilon = 0.74$ and (d) corresponds to $\epsilon = 0.98$. The phase space-filling nature of the dynamics does not change with increasing the value of $\epsilon$.}
\end{figure*}

\begin{figure*}
\includegraphics[scale=0.40]{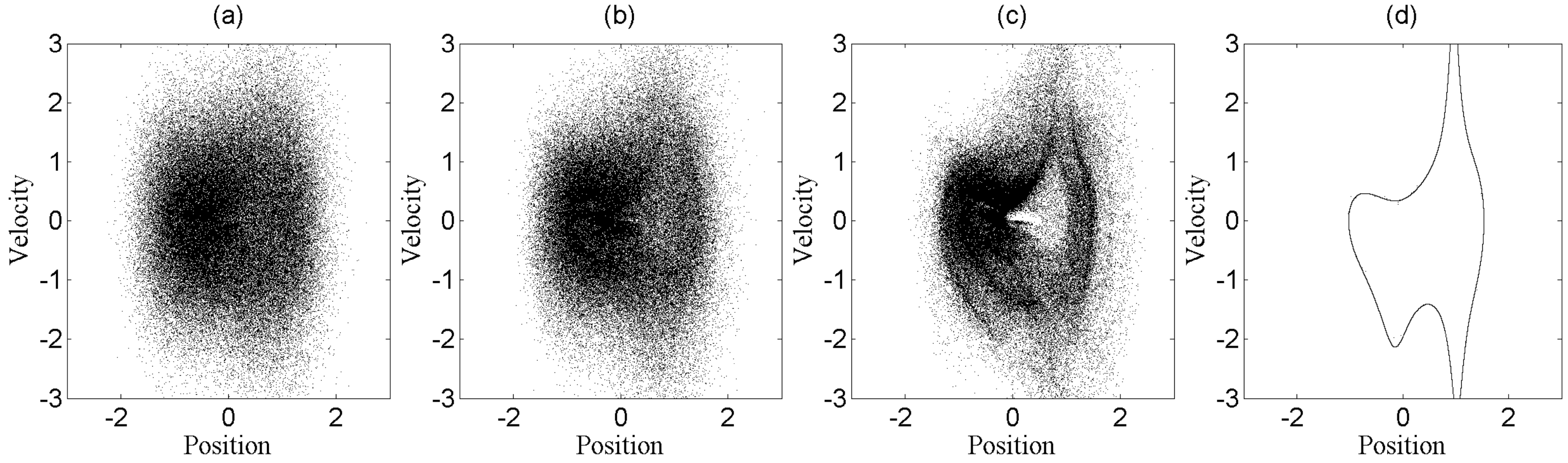}
\caption{\label{fig:FIGURE2} Phase space trajectory for Hoover-Holian dynamics. (a) corresponds to $\epsilon = 0.25$; (b) corresponds to $\epsilon = 0.50$; (c) corresponds to $\epsilon = 0.74$ and (d) corresponds to $\epsilon = 0.98$. Both position and velocity are initialized at 1. The nature of dynamics changes from being phase space-filling to a limit cycle through multifractals as $\epsilon$ increases.}
\end{figure*}

The two  thermostats result in different dynamics: LT produces a phase space-filling dynamics even under large values of $\epsilon$ \cite{hoover_posch} as shown in Figure \ref{fig:FIGURE1} while HH results in a dynamics that show a gradual shift from being phase space-filling to a limit cycle through intermediate multifractal dynamics \cite{sprott_et_al} as shown in Figure \ref{fig:FIGURE2}. 

We would like to stress that LT equations of motion cannot correctly simulate the non-equilibrium process \cite{leimkuhler}, but nevertheless we use LT to show that MaxRent can approximate $\rho_t(\Gamma)$  both for phase space-filling as well as multifractal dynamics. 

\begin{figure*}
\includegraphics[scale=0.425]{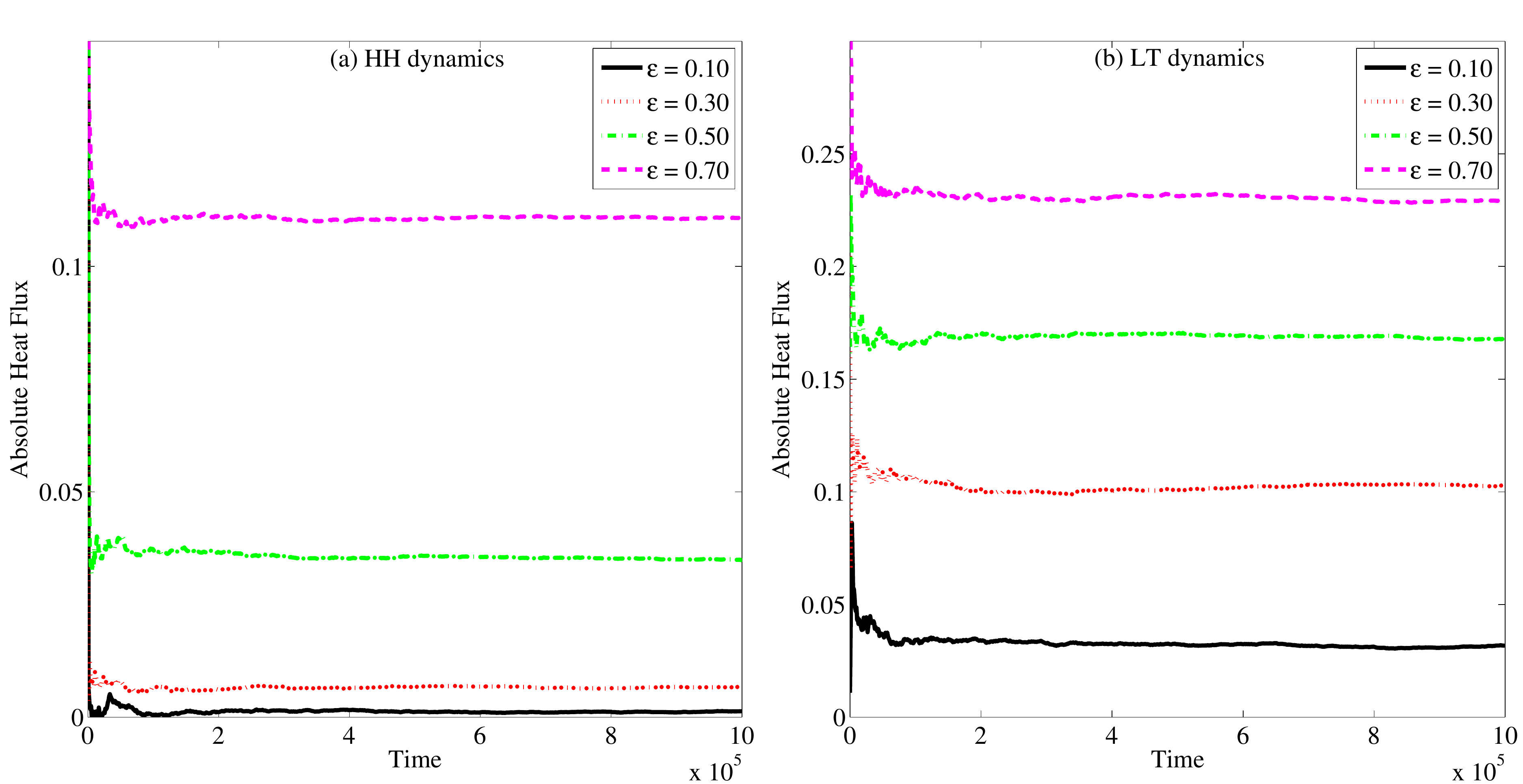}
\caption{\label{fig:FIGURE3} Check for steady state condition using four different values of the protocol $\epsilon$: (a) for HH dynamics and (b) for LT dynamics. In steady state, there is no appreciable change in the value of heat-flux with time. The figures indicate that the system reaches a steady state under the imposed condition fairly early in the simulation.}
\end{figure*}

It is interesting to see that in both the thermostats considered, the energy current is independent of the configurational variables (\ref{energy_current}). Consequently, there is no contribution of potential energy transport towards heat current.
\begin{equation}
\begin{array}{cl}
\dot{E} = & \dfrac{d}{dt}\left[ \dfrac{1}{4}x^4 + \dfrac{1}{2}p^2 \right] = x^3\dot{x} + p\dot{p} \\
& = \phi p^2 + \psi p \ \ \text{for LT} \\
& = \eta p^2 + \xi p^3 \ \ \text{for HH}
\end{array}
\label{energy_current}
\end{equation}
Since we limit the focus of this study to NESS and we make use of the ergodic hypothesis later on to estimate the density function from a single time trajectory, it is necessary to ascertain if these thermostatted dynamics can produce steady state conditions under the imposed temperature gradient. Figure \ref{fig:FIGURE3} plots the absolute value of heat flux for both the HH as well as LT dynamics for four different values of the protocol. It can be clearly seen that the system reaches steady state very soon after the nonequilibrium conditions are imposed. Throughout the rest of the paper, we assume that the steady state has set in so that $\partial \rho / \partial t = 0$.

\section{\label{section3}Principle of Maximum Relative Entropy \protect\\}
Given a set of constraints $\langle F_j \rangle$ for the present state of a system and a previously known state of the same system $\rho'(\Gamma)$, the principle of maximum relative entropy (MaxRent) states that the probability density $\bar{\rho(\Gamma)}$ that best approximates $\rho_t(\Gamma)$ can be obtained by maximizing the relative entropy subject to the constraints on the average value. The mathematical foundations of MaxRent can be found in literature \cite{shore_johnson, shore_johnson2, caticha_maxrent, caticha_maxrent3, giffin_thesis}. It has been argued that MaxRent is the only general tool for updating probabilities \cite{caticha_maxrent2, banavar_maxrent}. In conjunction with the constraints, their corresponding Lagrange multipliers and the known state of the system, the entropy functional to be maximized for MaxRent becomes:
\begin{equation}
\Delta H \equiv -k_B \rho(\Gamma) \log \left[ \dfrac{\rho(\Gamma)}{\rho'(\Gamma)} \right] - \sum\limits_{j=1}^{N}\lambda_j \left[\int\limits_{\Omega}F_j(\Gamma)\rho(\Gamma)d\Gamma - \langle F_j \rangle \right].
\label{maxrent_entropy}
\end{equation}
The first term is the Kullback-Leibler divergence \cite{banavar_maxrent, marc_pep} that measures the discrepancy between the distributions $\rho$ and $\rho'$, and is always $\geq 0$, with the equality holding only if $\rho = \rho'$ everywhere except, perhaps, at the points that form a zero measure set. The MaxRent approximated distribution function, $\bar{\rho(\Gamma)}$, can be obtained by taking variation of (10) with respect to the unknown distribution function $\rho(\Gamma)$, 
\begin{equation}
\bar{\rho }\left( \Gamma  \right)=\dfrac{1}{Z}\rho '\left( \Gamma  \right){{\exp}\left[{-\sum\limits_{j}{{{\lambda }_{j}}{{F}_{j}}\left( \Gamma  \right)}} \right] }.
\label{maxrent_approx_dist}
\end{equation}
One needs to find the unknown Lagrange multipliers, as in the MaxEnt method, to completely identify the approximated distribution function. In general, $\rho'(\Gamma)$ represents a known distribution which in this case pertains to a known detailed state of the system. The optimized distribution function $\bar{\rho}(\Gamma)$ represents the best “guess” at the true distribution function, $\rho(\Gamma)$, while simultaneously satisfying the constraints $\langle F_j \rangle$ and differing as little as possible from the known state of the system. When there is no prior information available, it makes sense to pre-assign equal probability to each state (i.e. a uniform density function) and the formulation boils down to MaxEnt (\ref{maxent_approx_dist}). Thus, the difference between MaxRent and MaxEnt comes from the additional ability of the former to handle prior information of any type. The success of MaxRent lies in careful selection of the prior distribution and identification of constraints not captured in the reference distribution \cite{banavar_maxrent_application}. In fact, it has recently been argued that equilibrium statistics can be inferred much more easily by using MaxRent than MaxEnt \cite{marc_pep}. 

For the problem in hand, i.e. to approximate the entire spectrum to nonequilibrium steady state distributions we propose to use MaxRent in the following manner. Instead of choosing an equilibrium prior distribution function, we select a nonequilibrium prior distribution function from amongst the states whose information is known to us in detail. To do so we assume that information for four steady states (out of the 80 possible) corresponding to $\epsilon = 0.20, 0.40, 0.60$ and $0.80$ is known to us in detail. The detailed information is transcribed into a probability distribution which can then be used as a prior distribution function. In this study, we use the probability distribution function calculated at a spatial resolution of 0.05 using 100,000 position-velocity data points (obtained from the simulations) as the prior distribution function. For the remaining 76 states, we obtain simple constraints that bear a direct physical meaning. While finding the distribution function of these states, we use the prior distribution function obtained from the protocol that lies the closest to the desired state. In essence, the method is similar to bootstrapping. Next we detail the constraints on the system.

\subsection{Identifying Constraints for NESS thermal conduction}
The performance of MaxRent depends on the choice of constraints, which importantly is problem-dependent. For example, in (\ref{energy_current}) the functional form of energy current is different for the LT and HH thermostats. Theoretically, it is possible to recover the \textit{exact distribution} if all joint moments of x and p i.e. $\langle x^mp^n \rangle$ for $m,n \geq 0$ are known. However, it is not possible to obtain such information from either simulations or experiments. Only a few moments are generally available. Additionally, higher order moments are usually associated with large errors and take a substantial time to converge. Therefore, in this study we have used only those constraints that are intuitive and have an order less than six. Such constraints bear a direct physical relevance and can be measured in laboratory settings easily. 

The simplest constraint obviously is the normalization property for density functions, i.e. $\int \limits_{\Omega} \rho(\Gamma) d\Gamma = 1$ . The other constraints are listed below.

\textit{No net mass current}: Absence of net mass current in this system i.e. average velocity of the system is zero:
\begin{equation}
\left\langle p \right\rangle =\int\limits_{\Omega }{p\rho \left( x,p \right)dxdp}=0
\label{no_mass_current}
\end{equation}

\textit{Fixed averaged position}: An implication of (\ref{no_mass_current}) is that the averaged position of the system must be time invariant and must equal some constant. Depending on the steady-state position of the system, this constraint can be written as:
\begin{equation}
\left\langle x \right\rangle =\int\limits_{\Omega }{x\rho \left( x,p \right)dxdp}={{x}_{0}}
\label{fix_position}
\end{equation}

\textit{Heat flux and flow of potential energy}: A set of constraints can be obtained from the definition of steady state systems: macroscopic properties of the system relax to fixed average values \cite{evans_book}. In the present context, the relevant macroscopic property is the heat current, which can be decomposed into two parts \cite{maruyama_md, sen_etal}: (i) transport of kinetic energy, and (ii) transport of potential energy. As has been shown before, the energy current is not dependent on configurational variables in this problem (see (\ref{energy_current})), and consequently, the rate at which kinetic energy is transferred is a proxy for heat flux and is a valid choice of constraint. In fact, in conventional molecular dynamics, (\ref{heat_flux}) represents the “heat flux”.
\begin{equation}
\int\limits_{\Omega }{\frac{{{p}^{3}}}{2}\rho \left( x,p \right)dxdp}=k_{0}
\label{heat_flux}
\end{equation}
Since the contribution of configurational variables towards energy current is absent, the averaged rate at which potential energy is transported from left to right (or vice versa) must be zero, and therefore,
\begin{equation}
\int\limits_{\Omega }{\frac{{{x}^{4}}p}{4}\rho \left( x,p \right)dxdp}=0
\label{pe_transport}
\end{equation}

\textit{Higher order derivatives of steady state quantities}: One can once again invoke the definition of steady state systems to argue that if (\ref{heat_flux}) is time invariant, then its subsequent time derivatives must be zero i.e.
\begin{align}
  & HH:\int\limits_{\Omega }{\left( -{{p}^{2}}{{x}^{3}}-\eta {{p}^{3}}-\xi {{p}^{5}} \right)\rho \left( x,p \right)dxdp}=0 \nonumber \\ 
 & LT:\int\limits_{\Omega }{\left( -{{p}^{2}}{{x}^{3}}-\phi {{p}^{3}}+\psi {{p}^{2}} \right)\rho \left( x,p \right)dxdp}=0
\label{hot_heat_flux}
\end{align}
Once can, simplify (\ref{hot_heat_flux}), by recognizing that $\phi$ is a constant (= 0.5) and $\psi$ is independent of $p$, with its average being equal to zero. Many subsequent time derivatives can also be equated to be zero. However, we do not consider these constraints because they are usually associated with substantial errors and are slow to converge. 

\textit{Averaged work done by the system}: Another constraint can be obtained by recognizing that in the steady state averaged work done by the external tethering potential must be time invariant i.e.
\begin{equation}
\left\langle {{x}^{4}} \right\rangle =\int\limits_{\Omega }{{{x}^{4}}\rho \left( x,p \right)dxdp}={{w}_{0}}
\label{avg_work}
\end{equation}

\textit{Energy Constraint}: So, far we have only chosen those constraints that can be measured easily in laboratory settings. We now place a constraint on the spatial variation of energy:
\begin{equation}
E\left( x \right)=\int\limits_{\Omega }{H\left( q,p \right)\delta \left( q-x \right)\rho \left( q,p \right)dpdx}
\label{energy_constraint}
\end{equation}
(\ref{energy_constraint}) is the limiting case for $E\left( a\le x\le b \right)$ as $a \to b$.  In the next subsection, we show that this constraint is associated with a Lagrange multiplier, $\beta(x)$, whose functional form is given by $1/T(x)$. This constraint is essential for MaxEnt to give sensible results even though local thermodynamic equilibrium (LTE) may not be valid. For MaxRent, although not mandatory, this constraint is imposed only for HH dynamics since LT does not sample the dynamics correctly. 

\subsection{Solving for the Lagrange multipliers}
In this section we obtain the functional form of the Lagrange multiplier $\beta(x)$ for both the MaxEnt and MaxRent methods. We assume that the thermodynamic temperature is applicable locally:
\begin{equation}
\frac{1}{T\left( x \right)}={{\left[ \frac{\delta S}{\delta E} \right]}_{x}}
\label{thermodynamic_temperature}
\end{equation}
where, $S$ is the Gibbs’ entropy. The Gibbs' entropy corresponding to the MaxEnt approximated distribution (\ref{maxent_approx_dist}) becomes:
\begin{equation}
\begin{array}{ccl}
S\left[\bar{\rho}\right] &=& - \int \bar{\rho}\left(x,p\right)\ln\left(\bar{\rho}(x,p)\right)dxdp\\
&=& \int \bar{\rho}(x,p)\left(\sum\limits_j\lambda_jF_j+\beta(x)H(x,p)\right)dxdp\\
&=& \int \bar{\rho}(x,p)\left(\sum\limits_j\lambda_jF_j\right) dxdp + \int\limits_{x}{\int\limits_{\Omega }{\bar{\rho }\left( q,p \right)\left[ \beta \left( q \right)H\left( q,p \right) \right]\delta \left( q-x \right)dqdpdx}}\\
&=& \int \bar{\rho}(x,p)\left(\sum\limits_j\lambda_jF_j\right) dxdp + \int \beta(x)E(x)dx.
\label{Gibbs_entropy_MaxEnt}
\end{array}
\end{equation}
In (\ref{Gibbs_entropy_MaxEnt}), we have omitted the integration constant. Taking variation of (\ref{Gibbs_entropy_MaxEnt}) with respect to $E(x)$ we get:
\begin{equation}
\beta \left( x \right)=\frac{1}{T\left( x \right)}
\label{beta_temperature_relation}
\end{equation}
Now, let us consider the case where $\rho'$ is not uniform. We assume that $\rho'$  has the functional form of  $\rho '=\exp \left( \sum\limits_{i}{{{\lambda }_{\rho ',i}}{{Y}_{i}}\left( x,p \right)}-{{\beta }_{\rho '}}\left( x \right)H\left( x,p \right) \right)$  where $Y_i(x,p)$ is a polynomial function of $(x,p)$. Using the approximated distribution (\ref{maxrent_approx_dist}), the Gibbs' entropy becomes:
\begin{equation}
\begin{array}{ccl}
S(\bar{\rho}) & = & -\int\bar{\rho}\left(x,p\right) \ln \bar{\rho}\left(x,p\right)\\
& = & -\int\bar{\rho}\left(x,p\right) \left( \sum\limits_i\lambda_{\rho^{\prime},i}Y_i + \sum\limits_j \left(\lambda_j + \lambda_{\rho^\prime,j}\right)F_j + \left( \beta(x) + \beta_{\rho^\prime}(x)\right) H(x,p) \right)dxdp .
\end{array}
\label{Gibbs_entropy_MaxRent}
\end{equation}
The Lagrangian multipliers, $\lambda_{\rho',j}$ and $\beta_{\rho'}(x)$ correspond to the prior distribution. Proceeding like before, we can see that the relationship between $\beta(x)$ and $T(x)$ becomes:
\begin{equation}
\beta \left( x \right)=\frac{1}{T\left( x \right)}-{{\beta }_{\rho '}}\left( x \right)
\label{beta_temperature_relation2}
\end{equation}

The remaining unknown Lagrange multipliers, $\lambda_j$ are solved using the Newton-Raphson technique. All necessary integrals are performed numerically using the trapezoidal rule. Sampling was confined to the [-3,3] interval for both the position and velocity.

\section{\label{Results} Approximating the NESS distributions under an evolving protocol}
We now show that MaxRent can be used to approximate the entire spectrum of the nonequilibrium distribution functions for the quartic oscillator under position-dependent temperature field. We proceed with maximizing (\ref{maxrent_entropy}) subject to the constraints highlighted before. Out of the four detailed states, whichever lies closest to the present state is taken as the prior distribution function. In simple terms it means that if we want to approximate the distribution function for $\epsilon_0 = 0.49$, then we choose the state that is closest to $\epsilon_0$ i.e. $\epsilon = 0.40$ as the prior distribution function. The distributions at each value of the protocol $\epsilon$ are compared to the one obtained through the MaxEnt framework. 

\begin{figure*}
\includegraphics[scale=0.425]{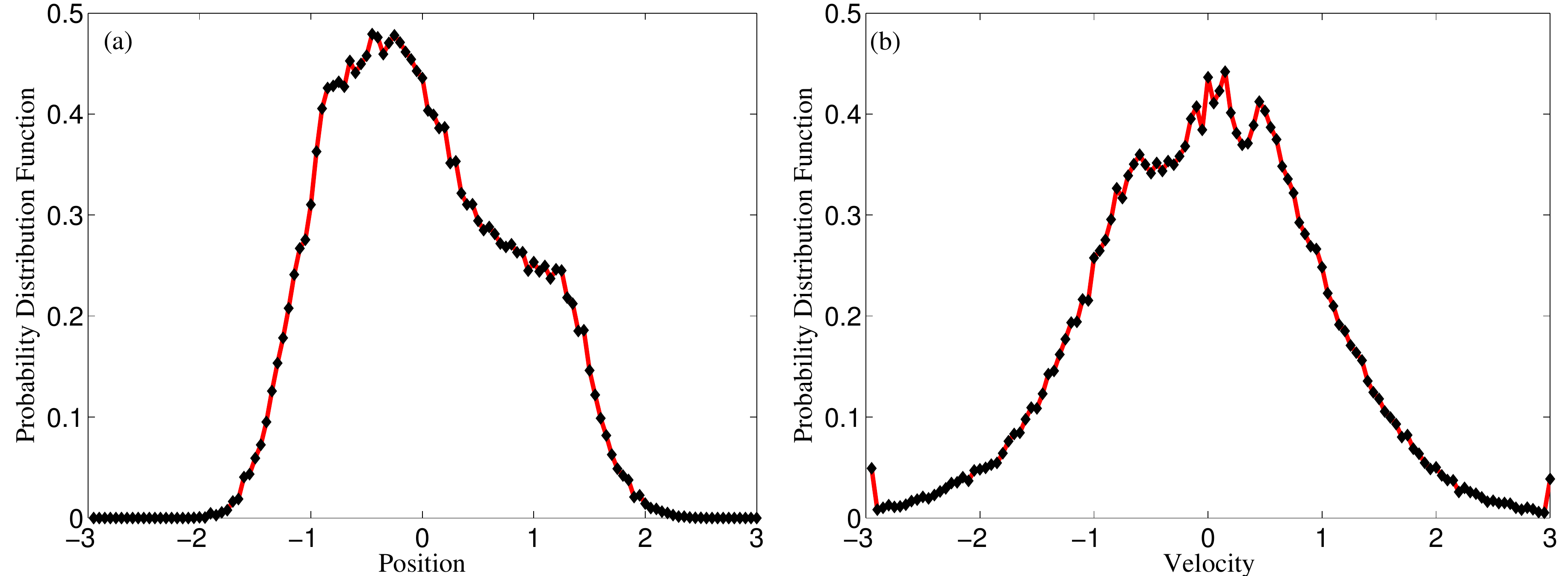}
\caption{\label{fig:FIGURE4} No effect of replacing $\langle \eta \rangle$ and $\langle \xi \rangle$ with arbitrary constraints in HH dynamics: (a) corresponds to the position distribution function and (b) corresponds to the velocity distribution function. Black diamonds indicate the case where $\langle \eta \rangle = \langle \xi \rangle = 1.0$. The red lines indicate the case where $\langle \eta \rangle = \langle \xi \rangle = 2.0$. Both the figures correspond to $\epsilon = 0.5$.}
\end{figure*}

For simplicity, in the case of HH thermostat we have assumed that the variables $\eta,\xi$ and $p$ are independent of each other (just like in equilibrium). Consequently, we replace $\langle \eta p^3 \rangle$ with $\langle \eta \rangle \langle p^3 \rangle$ and $\langle \xi p^5 \rangle$ with $\langle \xi \rangle \langle p^5 \rangle$, respectively for the constraint (\ref{hot_heat_flux}). Since in this work  we are not interested in the distribution of the thermostat variables, we replace their averages with numeric constants. This replacement does not alter the nature of position and velocity distributions drastically. Figure \ref{fig:FIGURE4} plots the approximated distributions obtained by replacing $\langle \eta \rangle = \langle \xi \rangle = 1.0$ and $\langle \eta \rangle = \langle \xi \rangle = 2.0$. Both the graphs overlap, suggesting that the distributions are independent of the constant value chosen. The reason may be attributed to the negligible contribution of the terms $\langle \eta p^3 \rangle$ and $\langle \xi p^5 \rangle$. For the rest of the analysis, we proceed with $\langle \eta \rangle = \langle \xi \rangle = 1.5$. 

\subsection{Marginal Position and Velocity Distributions}
\begin{figure*}
\includegraphics[scale=0.425]{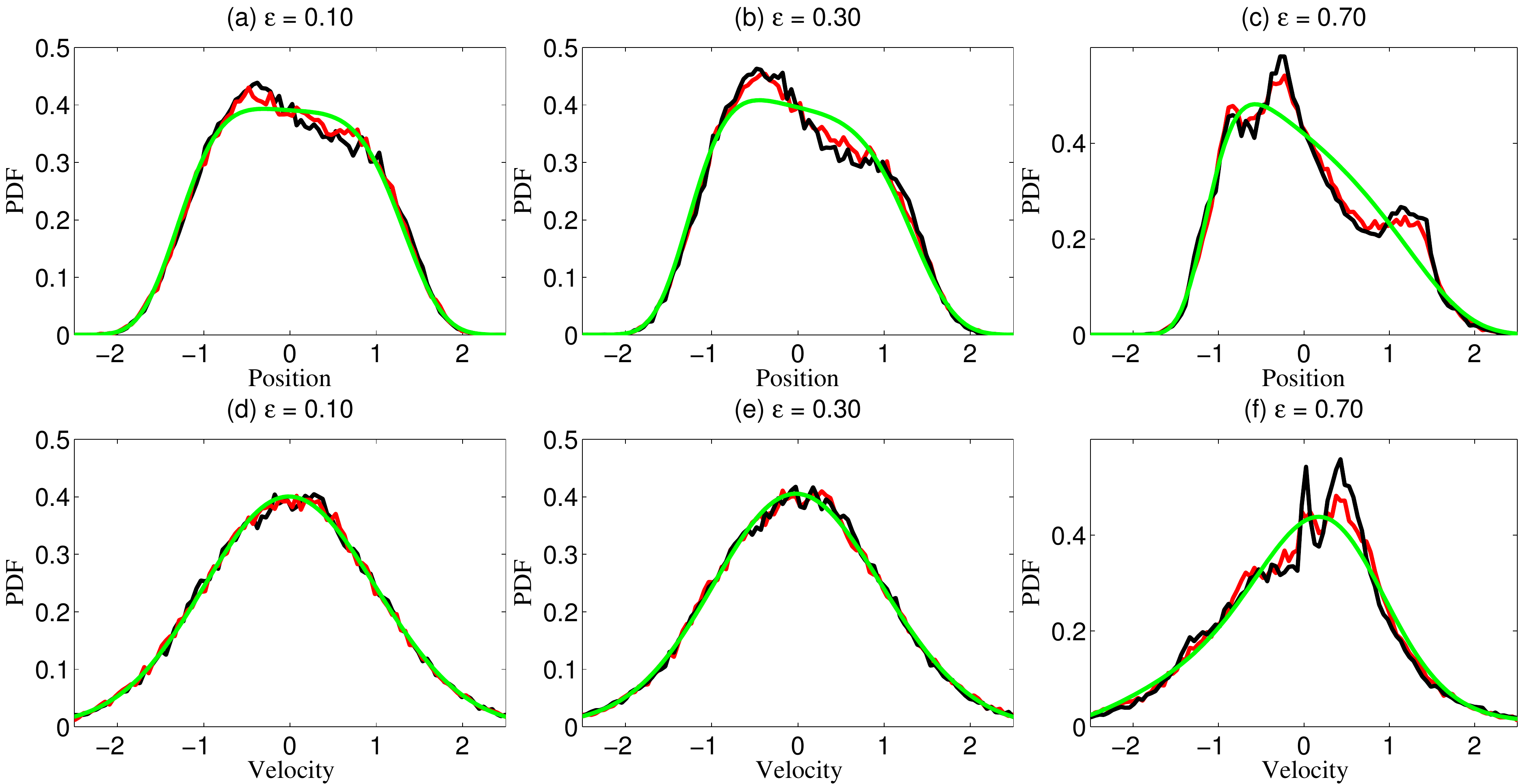}
\caption{\label{fig:FIGURE5} True (black), MaxRent (red) and MaxEnt (green) distribution functions of position (top row) and velocity (bottom row) for HH dynamics. Figures (a) and (d) correspond to $\epsilon = 0.10$. Figures (b) and (e) correspond to $\epsilon = 0.30$, and figures (c) and (f) correspond to $\epsilon = 0.70$. MaxRent accurately approximates and preserves the important features of the true distribution function. MaxEnt always returns a smooth distribution.}
\end{figure*}

\begin{figure*}
\includegraphics[scale=0.425]{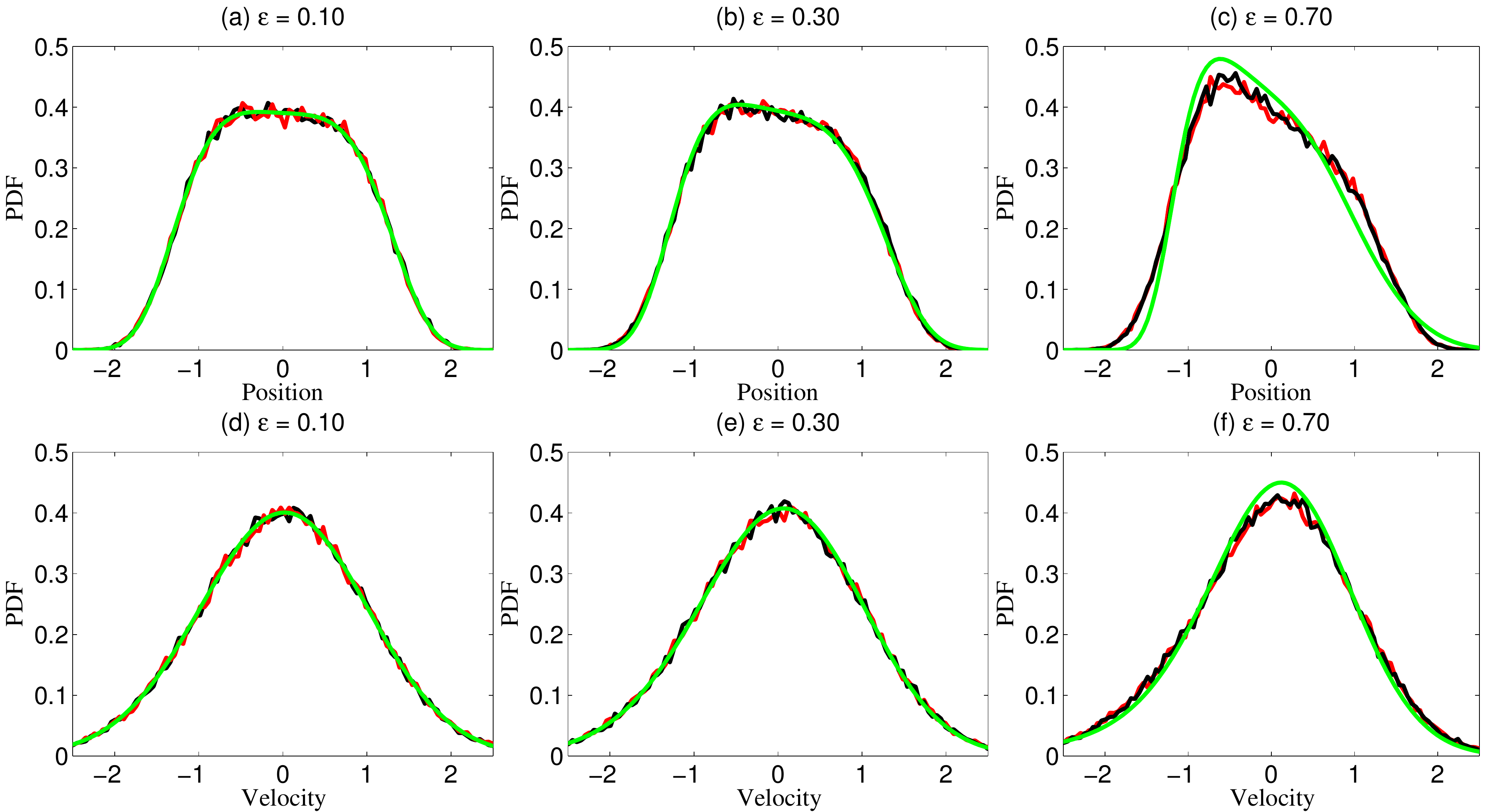}
\caption{\label{fig:FIGURE6} True (black), MaxRent (red) and MaxEnt (green) distribution functions of position (top row) and velocity (bottom row) for LT dynamics. Figures (a) and (d) correspond to $\epsilon = 0.10$. Figures (b) and (e) correspond to $\epsilon = 0.30$, and figures (c) and (f) correspond to $\epsilon = 0.70$. MaxRent accurately approximates and preserves the important features of the true distribution function.}
\end{figure*}

We begin by studying the ability of MaxRent to correctly describe the probability distributions of position and velocity (see Figures \ref{fig:FIGURE5} and \ref{fig:FIGURE6}) in three different regimes for both the thermostats: (i) near equilibrium (small temperature gradient with $\epsilon$ = 0.10), (ii) moderately away from equilibrium (moderately large temperature gradient with $\epsilon$ = 0.30) and (iii) far-from-equilibrium (large temperature gradient with $\epsilon$ = 0.70).

The black lines in Figures \ref{fig:FIGURE5} and \ref{fig:FIGURE6} indicate the true distributions, $\rho_t(x)$ and $\rho_t(p)$ (henceforth, referred to as $\rho_t(.)$), obtained directly from the simulation. The true position distribution function, $\rho_t(x)$, in Figure \ref{fig:FIGURE5} loses its symmetric nature even under a small temperature gradient. As $\epsilon$ increases, the asymmetry in  $\rho_t(x)$ also increases. It occurs because of the higher velocity of the oscillator when $x > 0$, due to higher temperature associated with this region. Consequently, the oscillator spends significantly less time in this region and hence, the asymmetry. 

At larger values of $\epsilon$, several kinks in $\rho_t(x)$ can be observed. These kinks are a characteristic of multifractal dynamics where some of the regions of phase-space are not visited (see Figure \ref{fig:FIGURE2} (c)). On the other hand, $\rho_t(p)$ in Figure \ref{fig:FIGURE5} are associated with negligible asymmetry until moderately large values of $\epsilon$. At larger $\epsilon$, however, the $\rho_t(p)$ shows the presence of kinks along with asymmetry. Any approximation to the true distribution functions must be able to capture these essential features. In Figure \ref{fig:FIGURE5}, we can clearly observe that MaxRent is able to approximate the true distribution functions faithfully for all strengths of nonlinearity. Under large temperature gradient (see Figure \ref{fig:FIGURE5} (b) and (c)), MaxRent accurately captures the peaks and the dips of the distribution as well as the asymmetry. On the other hand, MaxEnt washes out all the important features of $\rho_t(x)$ and $\rho_t(p)$, including the kinks. Its performance is rather poor when dealing with multifractal dynamics.
 
For LT dynamics (see Figure \ref{fig:FIGURE6}), there are no kinks owing to the phase space-filling dynamics. At larger values of $\epsilon$, we again observe asymmetry in the position distribution function. In this case as well, MaxRent performs better than MaxEnt in approximating $\rho_t(.)$. For HH dynamics (Figure \ref{fig:FIGURE5}), all the three distribution functions overlap at large absolute value of the independent variable. Interestingly, the same does not occur for LT dynamics. In Figure \ref{fig:FIGURE6} (c), there is a significant deviation between the distribution functions at large values of position. This is because of LT thermostat does not sample the imposed temperature accurately, and consequently, the use of energy constraint in MaxEnt leads to inconsistent distributions. At this stage we would like to point out that the performance of MaxEnt (as well as MaxRent) will improve if more constraints were chosen. 

\subsection{Quantifying the difference from the true distribution function }
\begin{figure*}
\includegraphics[scale=0.425]{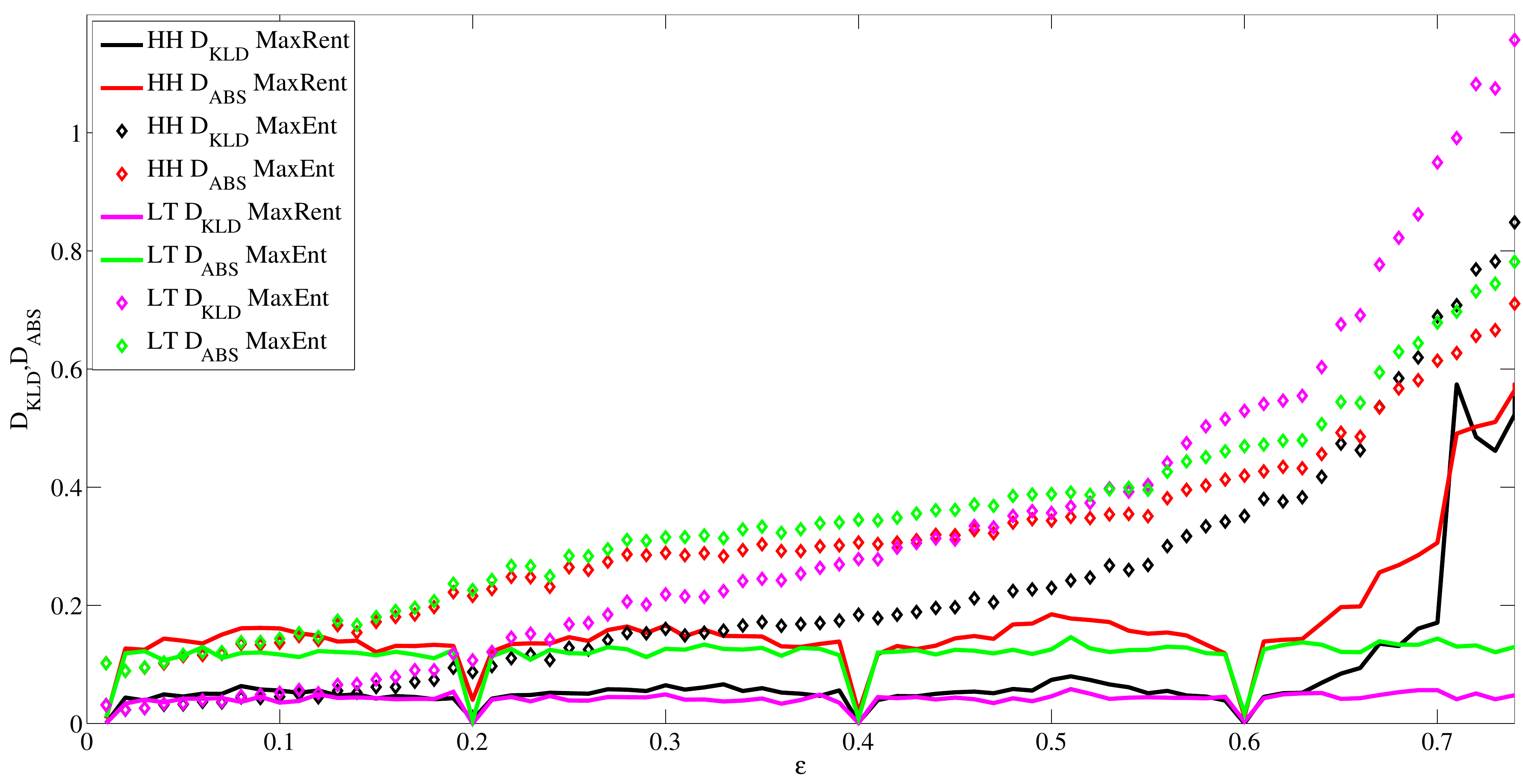}
\caption{\label{fig:FIGURE7} Comparison of Kullback-Leibler divergence (\ref{KLD}) and absolute error (\ref{D_ABS}) for the MaxRent and MaxEnt approximated distribution functions. Solid lines indicate the value of the metrics for MaxRent while dots indicate the values for MaxEnt. Apart from cases very close to equilibrium cases ($\epsilon \leq 0.04$), the deviation from true distribution is greater for MaxEnt than MaxRent. The deviation increases with increasing value of $\epsilon$ for MaxEnt, while it remains fairly constant over a large range of $\epsilon$ for MaxRent.}
\end{figure*}

The superiority of MaxRent becomes apparent when we calculate the difference between the approximated distributions and $\rho_t(\Gamma)$. We use two functions for this purpose – (i) Modified form of the Kullback-Leibler divergence,  with $\rho_t(i)$ being the true marginal distribution corresponding to the variable $i$ and $\bar{\rho_i}$  being the approximated marginal distribution corresponding to the variable $i$,
\begin{equation}
{{D}_{KLD}}\left( \bar{\rho} ||{{\rho }_{t}} \right) = \sum\limits_{x}{\bar{\rho} \left( x \right)\log \left[ \frac{ \bar{\rho} \left( x \right)}{{{\rho }_{t}}\left( x \right)} \right]} + \sum\limits_{p}{\bar{\rho} \left( p \right)\log \left[ \frac{ \bar{\rho} \left( p \right)}{{{\rho }_{t}}\left( p \right)} \right]}
\label{KLD}
\end{equation}
and, (ii) the absolute difference defined through:
\begin{equation}
{{D}_{ABS}}\left( \bar{\rho} || {{\rho }_{t}} \right)=\sqrt{\sum\limits_{x}{{{\left[ \bar{\rho} \left( x \right)-{{\rho }_{t}}\left( x \right) \right]}^{2}}} + \sum\limits_{p}{{{\left[ \bar{\rho} \left( p \right)-{{\rho }_{t}}\left( p \right) \right]}^{2}}}}
\label{D_ABS}
\end{equation}
The comparison is shown in Figure \ref{fig:FIGURE7}. The deviation of MaxRent approximated distributions from the true distributions is quite consistent for the better part of the spectrum. At larger gradients, though, the difference increases, indicating that the limited number of constraints chosen is not sufficient to accurately approximate the true distribution. Nevertheless, the results suggest that if slight errors are permissible, MaxRent can be used to approximate the unknown true distributions with limited data with better accuracy than MaxEnt.

\subsection{Effect of changing the prior distribution}
\begin{figure*}
\includegraphics[scale=0.425]{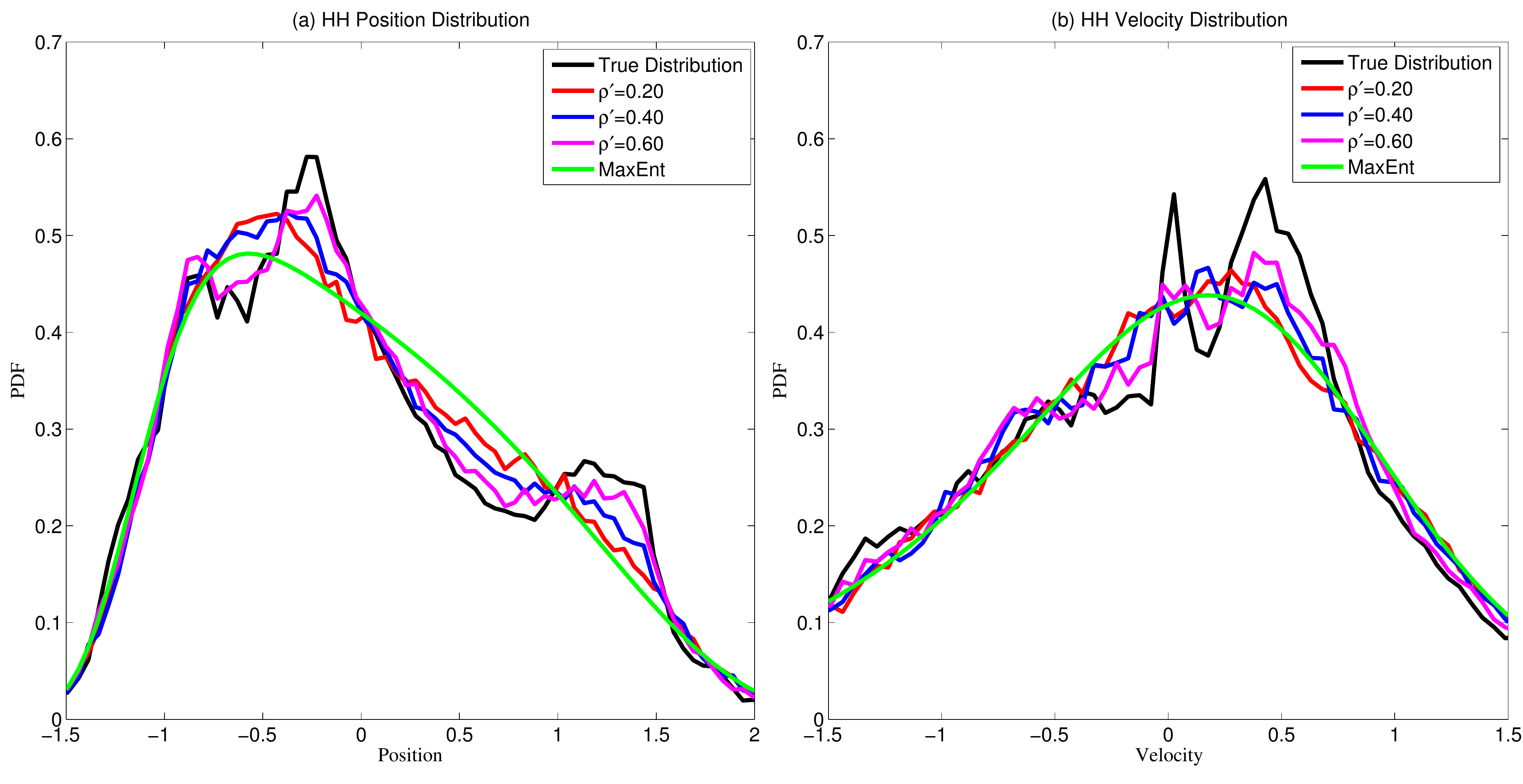}
\caption{\label{fig:FIGURE8} Distribution functions for HH dynamics with external protocol $\epsilon = 0.70$ using different prior distributions: (a) marginal distribution of position, and (b) marginal distribution of velocity. MaxRent results in a better approximation than MaxEnt for all the cases considered. But, the performance of MaxRent decreases as the prior distribution gets further away from the unknown true distribution.}
\end{figure*}

\begin{figure*}
\includegraphics[scale=0.425]{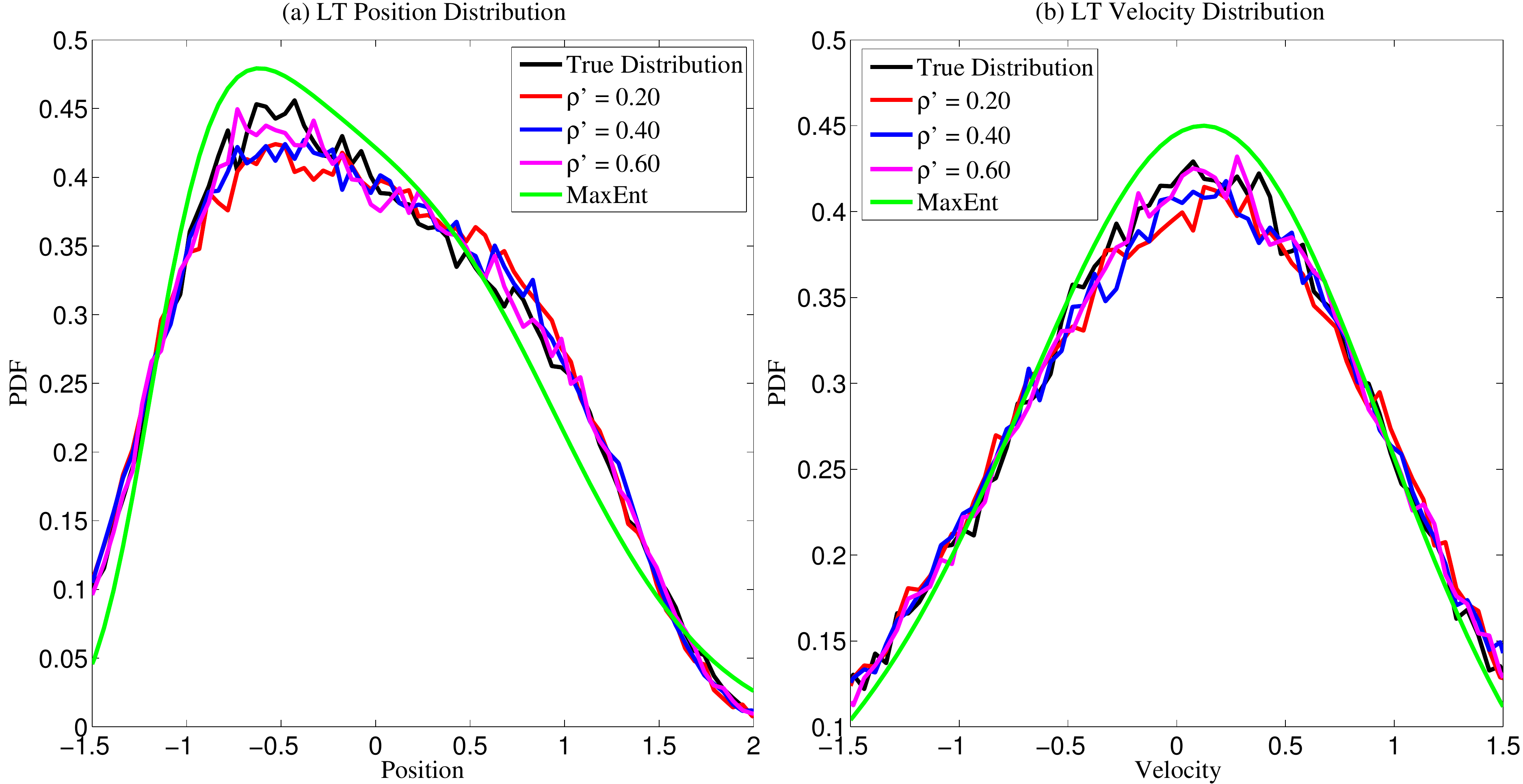}
\caption{\label{fig:FIGURE9} Distribution functions for LT dynamics with external protocol $\epsilon = 0.70$ using different prior distributions: (a) marginal distribution of position, and (b) marginal distribution of velocity. MaxRent results in a better approximation than MaxEnt for all the cases considered. MaxEnt approximation is substantially far from the true distributions at larger values of $x$. }
\end{figure*}

So far, we have used the closest amongst the four known states for calculating the approximated distribution function. It could be argued that the efficacy of MaxRent is due to selecting a prior distribution close to the unknown true distribution. Selection of a good prior distribution is crucial for the performance of MaxRent. In this section, we study how the approximated distributions change by changing the prior distribution. The effect is illustrated at $\epsilon = 0.70$. The resulting approximated distributions for the HH and LT thermostatted dynamics are shown in Figures \ref{fig:FIGURE8} and \ref{fig:FIGURE9}, respectively. 


\begin{figure*}
\includegraphics[scale=0.425]{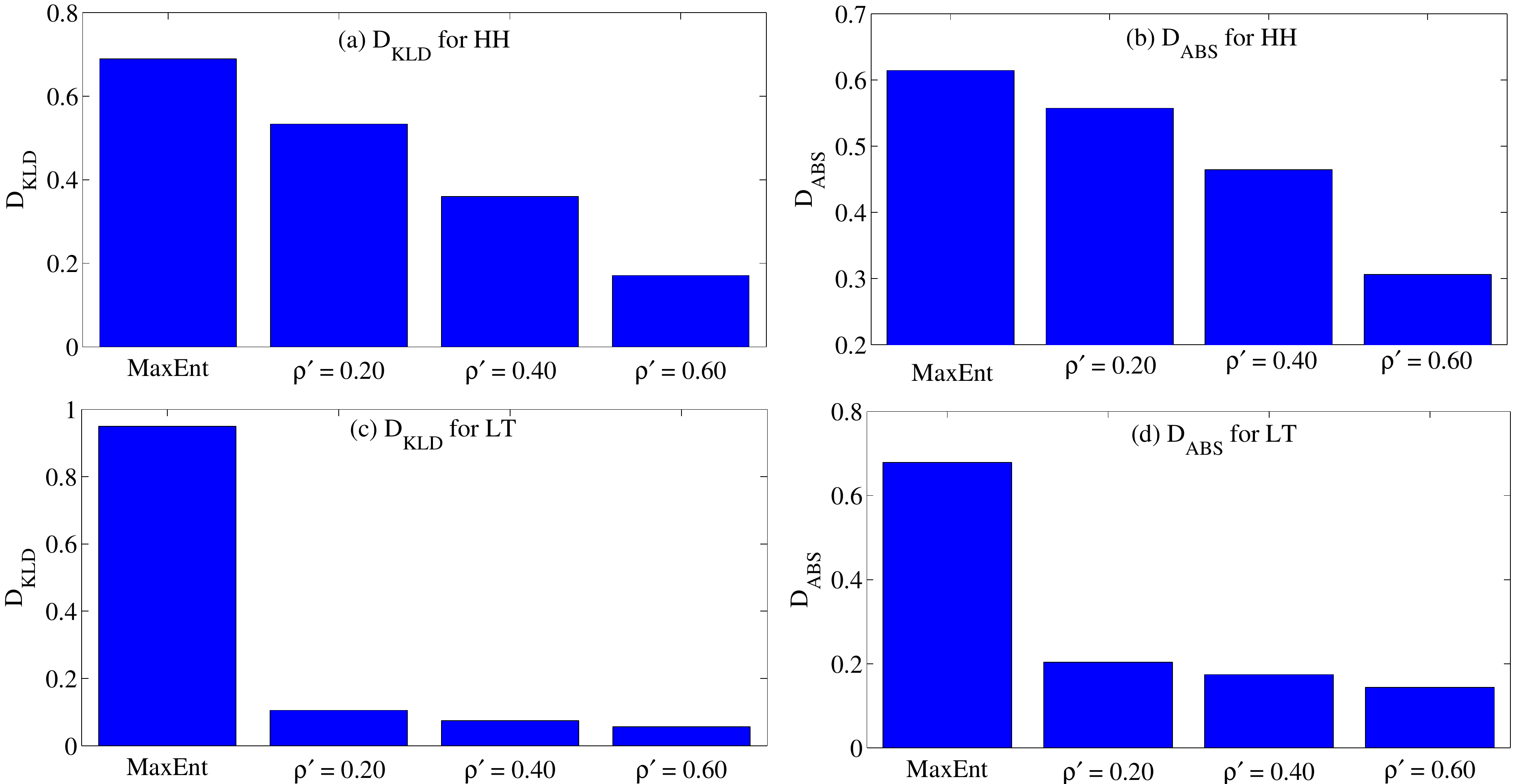}
\caption{\label{fig:FIGURE10} Comparison of (left) $D_{KLD}$ and (right) $D_{ABS}$ for different prior distributions due to (top row) HH dynamics and (bottom row) LT dynamics with $\epsilon = 0.70$. The closer the bars are to zero, the better the approximation to the true distribution function. MaxEnt results in the worst approximation.}
\end{figure*}

As expected, for HH dynamics, the closer the prior distribution is to $\rho_t(\Gamma)$, the better the approximation due to MaxRent. However, even with prior distributions as far away as $\epsilon = 0.20$, MaxRent results in a better approximation than MaxEnt and is able to capture the features of the true distribution. A similar behavior was observed for LT dynamics as well, although the difference amongst the different approximated distributions was not too large to begin with. The approximated distributions due to the different prior distributions (for both HH and LT) are compared in Figure \ref{fig:FIGURE10} using the two functions, (\ref{KLD}) and (\ref{D_ABS}). The error due to MaxRent decreases monotonically as the protocol comes closer to the current state.

\section{\label{Conclusion}Conclusions\protect\\}

In this work, we show through numerical examples that in a protocol controlled NESS it is not mandatory to perform detailed experiments for every value of the protocol to obtain statistics for the entire range of nonlinearity. We demonstrate this procedure for the toy model of a $\phi^4$ chain comprising a single lattice point. The system is brought out of equilibrium through a protocol-driven and position-dependent temperature field. This temperature field is imposed using two different thermostats. Using limited data in the form of simple constraints, and detailed description (obtained from simulations) of few of the states is sufficient to approximate the distribution functions for the entire range of nonlinearity. To do so, we maximize the relative entropy functional, (\ref{maxrent_entropy}), with the method of Lagrange multipliers using the Newton-Raphson technique. The detailed description is assumed to be given in form of the distribution function for four out of the 80 states that the protocol explores. 

The MaxRent approximated distribution functions are compared with those of the true and MaxEnt approximated distribution functions. We observe that MaxRent represents a marked improvement over the MaxEnt framework. For smaller values of $\epsilon$, both the MaxRent and MaxEnt perform comparably. However, at larger $\epsilon$, MaxRent approximates the true distributions in a much better way. In the absence of any prior information available, MaxRent defaults to MaxEnt. For the problem in hand, we observe that if MaxEnt were to perform comparably with MaxRent, then many more constraints would be needed, implying that one must perform detailed experiments for every value of the protocol. 

MaxRent has the ability to reproduce the true distribution in an accurate way because of the additional information that gets incorporated through the prior distribution function. However, its choice is subjective. The MaxRent approximated distribution functions have a remarkable dependence on the choice of the prior distribution. Likewise, the choice of constraints also impacts the performance of MaxRent. One must be careful to use appropriate constraints and prior distributions with MaxRent. Lastly, we would like to stress that although the method highlighted in this paper is for the simple case of a quartic oscillator, we believe that the method is useful in other non-equilibrium cases and would ease the life of the experimentalists, who now need not perform costly experiments for marginal changes in the external agent. 

\section{\label{Acknowledgement}Acknowledgement\protect\\}
The authors would like to thank Prof. William G. Hoover for providing stimulating ideas and critically appraising the first draft of this work.

\nocite{*}
\bibliography{apssamp}

\end{document}